RESEARCH ARTICLE                                                                                     OPEN ACCESS

# Multi-Lingual Ontology Server (MOS) For Discovering Web Services

Abdelrahman Abbas Ibrahim [1], Dr. Nael Salman [2]
Department of Software Engineering [1]
Sudan University for Science and Technology
Department of Computer engineering [2]
Palestine Technical University

**ABSTRACT**
Searching for appropriate web services on the internet is becoming more and more laborious, because it depends on human processing and evaluating of the available web services in UDDI repositories. Furthermore, if the requester language is different form available WSDL files then it would be more complicated. If this process could be done automatically, this will save effort and time. In order to make this factual, ontologies and semantic web technologies were used, ontology is needed to facilitate interoperability between agents and web services, to make them interoperate semantically, and to make processing of the data could be achieved automatically. In paper we propose an ontology server expected to help searching and selecting appropriate web service even if it's available in UDDI in different language.
***Keywords:-*** Ontology server, web service, ontology

## I.  INTRODUCTION

Service oriented architecture SOA and web services are new technologies for developing and building new generation of software. They compose up web service interfaces to make the whole system. However, Searching for appropriate semantic web services on the internet is becoming more and more laborious, because it depends on human processing and evaluating of available web services in UDDI repositories. Furthermore, if the requester language defer from available services then searching process will be more complicated and too difficult. If this process could be done automatically, this well saves effort and time. In order to make this factual, ontologies and semantic web technologies are being used. Ontology is needed to facilitate interoperability between agents and web services, to make them interoperate semantically, and semantic web services to describe services make processing of the data could be achieved automatically, a new generation of web services is about to be arise is the semantic web services which compose current web services with semantic web technologies. MOS ontology server well expected to help searching and selecting appropriate web service even if it is available in different language, for instance the descriptive file of the services may found in WSDL, which uses English as basic language while the end user uses his native language to find out particular web service.

This paper is organized in five sections. In the first section, we covered web service technology, while the second discusses ontology and semantic web, the third section give a brief information about ontology server and its main requirements, the next section assigns a wide area for proposed ontology server and showing its main components and how these components work together to fulfill its whole tasks, and lastly there is the related work and the summary.

## II.  WEB SERVICES

Web services are new paradigm used to build up a software system by composing them using new software architecture called service oriented architecture (SOA), web services allow us to reuse and combine software components via standardized interfaces, arbitrarily combining them to serve our needs (Bin et al., 2005).

A web service could be developed and publishing description of its interfaces and functions that it provide in a file developed with web service description language (WSDL), (Haas and Brown, 2004) , this file could be stored in central repository called Universal Description discovery and integration (**UDDI**) (Christensen et al., 2001) , via this repository WS requester could search and find the specification and full description of needed web service, to use this WS, the requester should be bind to the WS provider and then use it.

There are many technologies related to Web service, which facilitated description, publishing, discovering and accessing for instance Simple Object Access Protocol (**SOAP**) facilitate the operation of connecting and exchanging of xml files via networks over the **HTTP** and similar protocols. **WSDL** This standard allows a service interface and its bindings to be defined. **UDDI** standard Defines the components of a service specification that may be used to discover the existence of a service.**WS-BPEL** Web Service – Business Process Execution Language A standard for workflow languages used to define service composition(Alonso et al., 2004).

**ONTOLOGY AND SEMANTIC WEB**

The major drawback of current markup languages such as HTML and XML, which are used frequently to develop web applications, is that their documents do not convey the meaning of the data contained in them; they only provide syntactical format and not semantics (Berners-Lee et al., 2001). The semantic web allows the representation and exchange of data in a meaningful way, facilitating automated processing in web documents. Semantic web is envisioned as the next generation of the current web, the next generation will expand upon the prowess of the current web by adding machine readable information and automated services.

According to (Gruber, 1993), ”*the explicit representation of the semantic underlying data, program, and other web resources will enable a knowledge based web that provides a qualitatively new level of services*.” Ontologies provides an explicit representation of the semantic – ontology will be discussed in details in the next few paragraph – the combination of ontologies with the web has the potential to overcome many of the problems in knowledge sharing and reuse in information integration (Fensel et al., 2006) (Gašević et al., 2009).

Ontology is "an explicit **specification** of **conceptualization" (Gruber, 1993)**, this is the most concise definition; conceptualization means an abstract, simplified view of the world. Every conceptualization based on concept, object, and other entities that are assumed to exist in an area of interest, and the relation that exist among them, specification means a formal and declarative representation which implies that an ontology should be machine-readable. Ontologies provide a number of features including facilitating interoperability.

## III. ONTOLOGY SERVER

As mentioned earlier in previous sections, the ontology is a complex information object consists of many entities in taxonomy with a complicated relationship between them. In order to manage this kind of information object we need appropriate information system which intended to manage large amount of information within an enterprise, this information system is the ontology Server. The ontology server is an information system that responsible for managing ontologies. This server provides some tools to achieve essential tasks such as developing and editing the ontology.

**Table 1: Ontology server requirements**

| Lifecycle Stage | Requirements |
|---|---|
| **At design-time** | An ontology server should provide tools such as editing tools to enable ontology engineers to enter, modify, and browse a developing ontology., certify an ontology, Manage Imported Ontology Modules, Abstract Data Types and Metaproperties, Version Control, Publishing, |
| **At commit-time** | A player wishing to join an exchange needs to commit to the ontology, integrating part of their local conceptual model with at least part of the ontology. It provides: Browsing Services, Find Relevant Fragments of the Ontology, Subscription Services, Multiple Natural Languages, . |
| **At run-time** | An ontology server can perform tasks like Maintain Directories of Players, Roles and Objects. Validate Messages, Broker |

services, and Archive Services

Ontology servers are closely related to Computer-Aided Software Engineering (CASE) tools, which are a relatively mature technology. CASE tools are generally used to support the design of a system. Ontology servers are mainly in three lifecycle stages; in each lifecycle it provides some tasks requirements. Table 1 shows these requirements according to lifecycle (Gaševic et al., 2009).

## IV. MULTI-LINGUAL ONTOLOGY SERVER (MOS)

Multi-Lingual Ontology Server (MOS) is an ontology server for discovering semantic web services. It will act as a mediator between Web Service Requester and the provider as well as UDDI repositories. This Ontology server proposed to be fitted in the web service architecture, the following figure shows MOS ontology server fitted in the WS architecture. The WS requester can send a request directly to WS registry, but in case of different language the request will go through the ontology server, the ontology server helps in two things, find appropriate web service and bind the requester with the web server's provider.

ontology server which intended to facilitate the process of finding appropriate web service not aware of whether the end user use the same language as WSDLs in UDDI. In the next section main components of the MOS will be presented in details.

### 4.1. LANGUAGE DETECTOR (L-DETECTOR)

The main purpose of the component is to detect requester software language, the MOS ontology server need to know the language that requester browser uses to search for a web service, in implementation phase L-detector could be outsourced from Google web service.

L-detector can verify inputted terminologies and key information of search according to the ontology portion of detected language. In addition, it can direct the ontology server to associated portion of ontology, which supports this language.

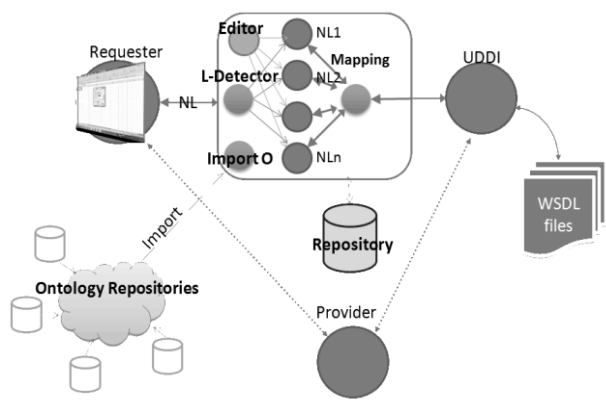

**Figure 2: MOS basic components**

### 4.2. ONTOLOGY IMPORTER (IMPORTO)

In case of language ontology portion is not available in MOS ontology server repository, the ontology server provides developer with a tool to search for the needed ontology portion of the ontology in public world ontology repositories and import it into MOS repository, **importO** software component achieve this task.

### 4.3. EDITOR

If the ontology portion is not available in MOS repository or in the international ontology

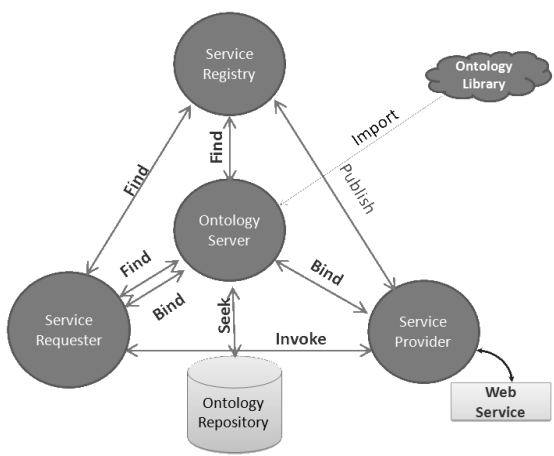

**Figure 1: Ontology server is fitted in-between**

## MOS SOFTWARE COMPONENTS

The MOS ontology server consist of many software components (web services) each component achieve appropriate task, figure (3) shows the main component of the Multi lingual repositories, then the ontology server provides a rich editor to help developers to design and edit the ontology portion manually, **Editor** Component is responsible for constructing ontologies at design time by adding classes and properties.

### 4.4. MOS REPOSITORY

Ontology repository is a persistent storage for ontology components, such as ontology metadata, ontology data, and any artifacts related to the ontology server, such as ontology portions, and supporting it with means for searching and indexing and more other facilities.

### 4.5. MAPPING SERVICE

Mapping software component responsible of mapping between ontology portion which holds the requester's search keywords in requester language and map it to UDDI WSDL specification language to facilitate searching in the original language, and translates the result back to requester language again.

### 4.6. ONTOLOGY PORTIONS

The ontology server has a repository containing the ontology which is divided into small portions each one represents specific language of ontology in the same domain, for instance Olympics ontology should have portions support English, Arabic, and French etc.

### 4.7. HOW MOS WORKS

The requester types keywords for searching in (UDDI interface) his own web browser or web service interface in order to find a specific or appropriate web service to be looking for in the public repositories UDDIs this request then is checked by the L-detect component which detects the original language of the requester after that L-detector matches the keywords with available ontology portions terms, if not available importO component will import related portion of the ontology from public ontology repositories, if it is available the Mapping component will map current keywords into WSDL language and resend back the result to requester who will make a decision to choose the suitable web services, then the ontology server will bind him to the provider.

## V. CASE STUDY

Suppose we have an UDDI repository containing a number of services, among them there are three services, the first has a WSDL file in Arabic, and the second in English and the last in French. They have the same functionality - finding the squire root of any number – French version has additional feature, it can find the squire root even the input number is negative or composite. And suppose we have a domain ontology consisting of three portions (aligned portions), supporting three languages Arabic, English and French, each term in Arabic portion has equivalents in both French and English. Now assume that the requester language is Arabic, and he looking for a service for finding squire root for numbers, no matter if this service described in Arabic or other languages, it is quite good if the service could find the squire root of negative numbers.

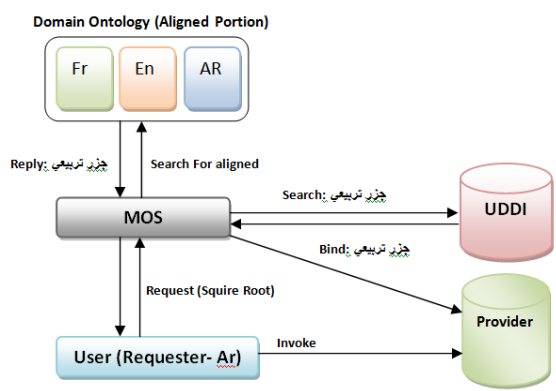

**Figure 3: Requesting Squire root service**

The requester will send a request to MOS which first detect the input language, the search for equivalent terms in ontology portions, then search for appropriate service in UDDI, If found it will bind it. In case of failing to find it, MOS will continue using other terms from the ontology. Figure .

## VI. RELATED WORKS

*Ling Wang* and *Jinli Cao (Wang and Cao, 2007)* proposed a Web services matchmaker that plays a role like an OWL-S/UDDI Matchmaker but is augmented with case-based reasoning, matching and ranking capabilities according to past experiences. *Xu Bin et al* (Bin et al., 2005) proposed a method to search web services according to domain ontology. Firstly, to get as much as possible WSDL files directly in the Internet, the WSDL crawler collects WSDL files from the web pages in XMethods1, Google2 and Baidu3. Secondly, to represent a travel domain, the travel ontology is given. Then domain vector is built according to the ontology. Thirdly, features are extracted from selected WSDL files to form the training set of SVM classifier. Experiment shows that the method is effective.

## VII. CONCLUSION

This paper proposes a framework for an otology server called Multi-lingual otology server (MOS) which provides requester with tools to help them in searching for web service in public repositories using their native languages without concerning with the language that the UDDI repository uses or provider's language. The ontology server contains L-detector, ontology Portion importer (ImportO), repository, Editor and other facilities.

## REFERENCE

[1] ALONSO, G., CASATI, F., KUNO, H. & MACHIRAJU, V. 2004. Web services, Springer.

[2] BERNERS-LEE, T., HENDLER, J. & LASSILA, O. 2001. The semantic web. Scientific american, 284, 28-37.

[3] BIN, X., YAN, W., PO, Z. & JUANZI, L. Year. Web services searching based on domain ontology. In: Service-Oriented System Engineering, 2005. SOSE 2005. IEEE International Workshop, 2005. IEEE, 51-55.

[4] CHRISTENSEN, E., CURBERA, F., MEREDITH, G. & WEERAWARANA, S. 2001. Web services description language (WSDL) 1.1.

[5] FENSEL, D., LAUSEN, H., POLLERES, A., DE BRUIJN, J., STOLLBERG, M., ROMAN, D. & DOMINGUE, J. 2006. Enabling semantic web services: the web service modeling ontology, Springer Science & Business Media.

[6] GAŠEVIC, D., DJURIC, D. & DEVEDŽIC, V. 2009. Model driven engineering and ontology development, Springer Science & Business Media.

[7] GRUBER, T. R. 1993. A translation approach to portable ontology specifications. Knowledge acquisition, 5, 199-220.

[8] HAAS, H. & BROWN, A. 2004. Web services glossary. W3C Working Group Note (11 February 2004), 9.

[9] WANG, L. & CAO, J. Year. Web Services Semantic Searching enhanced by Case Reasoning. In: Database and Expert Systems Applications, 2007. DEXA'07. 18th International Workshop on, 2007. IEEE, 565-569.